\documentclass[12pt,a4paper]{article}

\usepackage[utf8]{inputenc}
\usepackage{color}
\usepackage{parskip}
\usepackage{enumitem}
\usepackage{amsmath,amscd}

\usepackage{stmaryrd}
\usepackage{mathrsfs}
\usepackage{amsfonts}
\usepackage{amssymb,scalerel,stackengine}
\usepackage[most]{tcolorbox}
\newtcbox{\othermathbox}[1][]{nobeforeafter, math upper, tcbox raise base, 
	enhanced, rounded corners, colback=black!5, colframe=black}
\usepackage{empheq}
\usepackage[margin=1in]{geometry}
\usepackage{graphicx}
\usepackage{diagbox} 
\usepackage{multirow} 
\usepackage{array} 
\usepackage[leftcaption]{sidecap}
\usepackage{caption}
\usepackage{subcaption}
\usepackage{here}
\usepackage[colorlinks=true,allcolors=blue]{hyperref}
\usepackage{cite}
\let\OLDthebibliography\thebibliography
\renewcommand\thebibliography[1]{
	\OLDthebibliography{#1}
	\setlength{\parskip}{0pt}
	\setlength{\itemsep}{1.85pt plus 0.3ex}
}

\protected\def\verythinspace{%
  \ifmmode
    \mskip0.5\thinmuskip
  \else
    \ifhmode
      \kern0.08334em
    \fi
  \fi
}
\newcommand\beq{\begin{equation}}
	\newcommand\ee{\end{equation}}

\newcommand\cO{{\cal O}}

\def\barD{\overline{\text{D}}}
\newcommand\dd{\text{d}}
\newcommand\DD{\mathrm{D}}

\newcommand\ie{\textit{i.e.}}
\newcommand\eg{\textit{e.g.}}

\def\ee{{\rm e}}

\def\ax{{m=0}}
\def\ii{{\rm i}}
\def\a{A}
\def\b{B}
\def\c{C}
\def\d{D}

\def\hi{{\hat i}}
\def\hj{{\hat j}}

\def\ha{{\hat A}}
\def\hb{{\hat B}}

\def\hr{{\hat r}}
\def\mn{{\mu\nu}}

\def\de{\delta}

\def\pd{\partial}

\def\cd{\nabla}

\def\veps{\varepsilon}

\def\cY{\mathcal{Y}}

\def\WS{\hat{\Omega}_\text{(S)}}
\def\WH{\hat{\Omega}_\text{(H)}}

\def\barC{\overline{C}}

\def\tlC{\widetilde{C}}

\def\cO{\mathcal{O}}

\def\barm{{\overline{m}}}

\def\bareth{{\overline{\eth}}}

\def\barZ{{\overline{Z}}}

\def\th{\theta}

\def\tl{\widetilde}

\newcommand\Ylms[2]{{{}_{#1}{Y}^{#2}}}
\newcommand\barYlms[2]{{{}_{#1}{\overline{Y}}^{#2}}}

\newcommand\ThreeJ[6]{\begin{pmatrix}
			#1 & #3 & #5 \\
			#2 & #4 & #6
		\end{pmatrix}}
\newcommand\GG[3]{{{\cal G}_{#1}^{\, #2,\,#3}}}

\title{\bf Gyroscopic Gravitational Memory from quasi-circular binary systems}

\author{Guillaume Faye$^{1}$ and Ali Seraj$^{2}$ \vspace{1em}\\
	{\small $^{1}$ $\mathcal{G}\mathbb{R}\varepsilon \mathbb{C} \mathcal{O}$,  Institut d'Astrophysique de Paris, UMR 7095,}\\\vspace{.4em}
	{\small CNRS \& Sorbonne Universit{\'e}, 98\textsuperscript{bis} boulevard Arago, 75014 Paris, France}\\ 
	{\small $^{2}$ School of Physics, Institute for Research in Fundamental Sciences (IPM), }\\\vspace{-.2em}
	{\small P.O.Box 19395-5531, Tehran, Iran}}
\date{}
\begin{document}

\maketitle
Email: \href{mailto:faye@iap.fr}{faye@iap.fr} and 
\href{mailto:ali\_seraj@ipm.ir}{ali\_seraj@ipm.ir} \\

\begin{abstract}
Gravitational waves cause freely falling spinning objects to precess, resulting in a net orientation change called gyroscopic memory. In this paper, we will consider isolated gravitational sources in the post-Newtonian framework and compute the gyroscopic precession and memory at leading post-Newtonian (PN) orders. We compare two competing contributions: the spin memory and the nonlinear helicity flux. At the level of the precession rate, the former is a 2PN oscillatory effect, while the latter is a 4PN adiabatic effect. However, the gyroscopic memory involves a time integration, which enhances subleading adiabatic effects by the fifth power of the velocity of light, leading to a 1.5PN memory effect. We explicitly compute the leading effects for a quasi-circular binary system and obtain the angular dependence of the memory on the celestial sphere.
\end{abstract}

\newpage
\tableofcontents

\section{Introduction}

Spin and rotation effects play a crucial role in the dynamics of gravitating systems in general relativity (GR). The mass currents arising in rotating bodies, source a ``gravitomagnetic'' potential entering the parametrization of the linearized metric, with no Newtonian counterpart. The primary example of a stationary rotating solution is the Kerr black hole, which has many distinct features from a static black hole of the same mass. Examples of such features include the Lens-Thirring frame dragging~\cite{Misner:1973prb} and the black hole's ergosphere, whose interaction with the surrounding matter fields leads to interesting physical effects, such as  superradiance~\cite{Brito:2015oca} and the Blandford–Znajek process~\cite{Blandford:1977ds}. 

In addition to stationary situations, it is necessary to describe the
dynamical interaction between gravity and spinning matter. This topic has been
extensively studied, beginning with the seminal works of Mathisson and
Papapetrou~\cite{Mathisson:1937zz,papapetrou1951spinning}, who formulated the
motion and precession of test spinning particles in GR. Their results were
rephrased by Tulczyjew, and extended by Dixon and others to include the higher
multipole structure of test particles~\cite{tulczyjew1959equations, dixon1970dynamics, Dixon:1970zz, Dixon:1974xoz, 1977GReGr...8..197E}. The Mathisson, Papapetrou, Dixon (MPD) equations can be derived in full generality from an effective worldline action~\cite{Bailey:1975fe}, laying the basis of the effective field theory approach to gravitational waves (GW). This formalism may be used to describe self-gravitating systems of spinning particles in the post-Newtonian (PN) framework as an important class of GW sources (see~\cite{Tagoshi:2000zg, Faye:2006gx} for the derivation of the PN equations of motion for compact binaries from MPD equations, or~\cite{Porto:2016pyg} for the calculation of the corresponding Fokker Lagrangian, following theoretical grounds elucidated in~\cite{LS15b}). 

In this paper, we will be interested in spinning bodies as probes of GWs. In particular, we will investigate DC effects of GW that accumulate over time and lead to persistent observables lasting after the passage of the wave~\cite{Flanagan:2018yzh}. The primary example of such observables is the displacement effect, consisting of a net change in the distance between nearby freely falling test particles~\cite{Zeldovich:1974aa,Braginsky:1985vlg,braginsky1987gravitational,Christodoulou:1991cr,Blanchet:1992br}. If the test particle is endowed with a spin, as a gyroscope, an additional observable is the precession of the spin caused by the GWs. Different experiments to measure the effect can be thought of. One consists in computing the change in the spin of a rotating object with respect to a parallel-transported frame at finite distance, as suggested in~\cite{Flanagan:2018yzh}. Another observable is the precession of the spin of a gyroscope with respect to a comoving optical frame tied to distant stars, and the subsequent persistent rotation after the passage of the wave. The latter proposal was examined in~\cite{Seraj:2021rxd,Seraj:2022qyt}, where it was found that, restricting to the regime in which the spin vector is approximately parallel transported along a geodesic motion, GWs cause a precession in the local transverse plane of the propagation and is proportional to the inverse square of the distance to the source. As a result, the passage of the wave induces in that case a net rotation in the orientation of the gyroscope, dubbed as the ``gyroscopic memory''. The dynamics is however more involved when nonlinear spin interactions cannot be neglected so that the full MPD equations must be taken into account. In section~\ref{sec: Spin in GR}, we will clarify the limit in which the parallel-transport assumption is valid.

The goal of this paper is to investigate the PN expansion of the gyroscopic memory when the GWs are sourced by binary systems of compact objects. We will actually focus on \textit{quasi-circular} binaries, consisting of two non-spinning compact bodies on a nearly circular orbit whose radius adiabatically shrinks in time. We will address the following questions: (1) How does the test gyroscope's precession rate depend on its angular position on the celestial sphere? (2) What is the magnitude of the precession in terms of the parameters of the binary system? (3) What are the DC components of the precession and the accumulation effect resulting in the gyroscopic memory?

The paper is organized as follows. In section~\ref{sec: Spin in GR}, we introduce the MPD equations to show how, in a certain limit, the spin evolution reduces to the parallel transport used in deriving the gyroscopic memory effect in~\cite{Seraj:2021rxd,Seraj:2022qyt}. This construction and its implications are reviewed in~\ref{sec: formal}. In section~\ref{sec: multipole expansion}, we reformulate the precession rate and gyroscopic memory using ``spin-weighted'' functions on the celestial sphere. The latter approach is advantageous not only formally, due to the simplicity of the resulting expressions, but also for practical computations. Resorting to those tools, it is straightforward to perform the multipole expansion of the precession and to compute memory effects in terms of the radiative moments,  which is achieved in section~\ref{sec: multipole expansion}. In section~\ref{sec: PN}, we introduce quasi-circular binary systems, discuss their radiation, and compute the precession of a distant freely falling gyroscope as well as the gyroscopic memory in the PN framework. We conclude in section \ref{sec: conclusion}, by briefly discussing observational aspects of the gyroscopic memory.

\section{Brief review of spin dynamics in general relativity} \label{sec: Spin in GR}

\subsection{Effective approach to spinning objects}

The motion and precession of a free point-like body with velocity $v^\mu$ (normalized so that $v^\mu v_\mu = -c^2$), momentum $p_\mu$, carrying some spin represented by the antisymmetric tensor $S^\mn$ is given by the equations~\cite{dixon1970dynamics}
\begin{subequations} \label{eq:MPD}
\begin{align}
 \frac{\DD p_\mu}{\DD\tau} &= -\frac{1}{2} R_{\mn\rho \sigma} v^\nu S^{\rho \sigma}-\frac{c^2}{6} J^{\nu\lambda \rho \sigma} \cd_\mu R_{\nu\lambda \rho \sigma} + \cdots\,, \label{eq:MPD2}\\
 \frac{\DD S^\mn}{\DD\tau} &=2 p^{[\mu}v^{\nu]} +\frac{4 c^2}{3} R^{[\mu}{}_{\lambda \rho \sigma}J^{\nu]\lambda \rho \sigma}+\cdots\,, \label{eq:MPD1}
\end{align} 
\end{subequations}
with $\DD/\DD\tau \equiv v^\mu\cd_\mu$, where the Levi-Civita derivative $\nabla_\mu$, and the corresponding Riemann tensor $R^\mu_{~\nu\rho\sigma}$ are derived from the metric $g_{\mu\nu}$, and the square brackets denote the antisymmetrization over the enclosing indices. The tensor $S^\mn$ contains nonphysical degrees of freedom related to the choice of the body centroid. To fix them, the above equations must be supplemented by a spin supplementary condition~\cite{Corinaldesi:1951pb, 1979GReGr..11..149B, Costa:2014nta}. The covariant Tulczyjew condition $p_\nu S^\mn=0$~\cite{tulczyjew1959equations, Dixon:1964cjb, dixon1970dynamics} is particularly convenient, as it implies that in the ``center-of-mass'' frame --- defined as a frame whose temporal basis vector lies along $p^\nu$ ---, the spin vector is purely spatial. The internal structure of the body manifests itself through the Dixon moments, such as the quadrupole tensor $J^{\mn\rho\sigma}$, as well as higher-order moments, $J^{\nu\lambda\rho\sigma\tau}$, $J^{\nu\lambda\rho\sigma\tau\upsilon}$, etc., which enter the subleading corrections represented by the dots in~\eqref{eq:MPD}.

From general considerations in statistical physics, we expect that in thermodynamic equilibrium, the macroscopic state of a body--- and thus its Dixon multipole moments---should be entirely determined by its mass, energy, linear and angular momenta, equation of state, and stationary external gravitational field\footnote{Such equilibrium can only hold approximately, on timescales that are sufficiently short to prevent any significant exchange of energy with the environment so that, in particular, back-reaction effects remain negligible. See~section II.E in~Ref.~\cite{Diu85} for a discussion on macroscopic configurations of non-relativistic non-gravitating isolated systems at thermodynamic equilibrium. In binary systems, because the external field is not stationary, the best we can achieve is an ``instantaneous'' quasi-equilibrium, provided the relaxation time is much smaller than the orbital period. This practically implies that both companions should behave as fluids.}. This justifies to resort, when describing the dynamic of point-like particles, to effective Lagrangians that depend only on the velocities and spins of the particles, as well as the metric and (derivatives of) the Weyl tensor evaluated at the locations of the particles~\cite{LS15b, Marsat:2014xea}. Those Lagrangians can be used  to derive effective expressions for the Dixon moments. If tidal deformations are negligible compared to spin-induced ones, the Dixon quadrupole reads~\cite{Marsat:2014xea, Buonanno:2012rv}
\begin{align} \label{eq:J4}
J^{\mn\rho\sigma} &= \frac{3\kappa}{m c^4}v^{[\mu}S^{\nu]\lambda} S_\lambda{}^{[\rho} v^{\sigma]} \,,
\end{align}
where the dependence on the equation of state arises through the dimensionless constant~$\kappa$. Equations~\eqref{eq:MPD} and~\eqref{eq:J4} must be thought of as effective and, as such, they do remain valid for self-gravitating systems. It should be noted that only the first terms in the right-hand sides of each equation in~\eqref{eq:MPD} are universal, \ie, independent of the nature of the body, while the others terms containing Dixon moments depend on the system's equation of state. The full MPD equations for test extended bodies, in the zero radius limit, were first derived by Dixon~\cite{Dixon:1974xoz}, while Bailey and Israel~\cite{Bailey:1975fe} recovered this result starting from an effective Lagrangian (see \cite{Marsat:2014xea} for a short modern review). This approach was later incorporated in the effective field theory to investigate the dynamics of GW sources~\cite{Goldberger:2004jt,Porto:2016pyg}.

Contracting the momentum with~\eqref{eq:MPD1}, employing the Leibniz rule, the supplementary condition and~\eqref{eq:MPD2} imply that $	p_\nu \DD S^\mn/\DD\tau =\cO(S^2)\,.$ Using this information in~\eqref{eq:MPD1} next reveals that $2p_\nu p^{[\mu}v^{\nu]} = \cO(S^2)$. Defining the bare mass as $m\equiv -p^\mu v_\mu/c^2$, it follows that $p^\mu p_\mu + m^2 c^2 = \cO(S^2)$ and  $p^\mu =m v^\mu +\cO(S^2)\,.$
Inserting this back in~\eqref{eq:MPD1}, we find that 
\begin{align}
	\frac{\DD S^\mn}{\DD\tau} =0 + \cO(S^2)\,,
\end{align}
\ie, the spin obeys the parallel transport equation to leading order in the spin. Moreover, the body's worldline is geodesic up to linear corrections in the spin, namely, $\DD v^\mu/\DD\tau =\cO(S^\mn)$. In the rest of this paper, we will neglect higher order spin corrections and will therefore restrict our attention to the parallel transport along a geodesic. In this regime, it is more convenient to define a spin vector $S^\mu\equiv -\veps_{\mn\rho\sigma} p^\nu S^{\rho\sigma}/(2m c)$, so that, at leading order in the spin, the evolution equations reduce to 
\begin{align} \label{eq:parallel transport}
	\frac{\DD v^\mu}{\DD\tau} &= 0, \qquad \frac{\DD S^\mu}{\DD\tau} =0\,.
\end{align}

\subsection{Parallel transport in orthonormal frames}

The parallel transport equation can alternatively be formulated in a local orthonormal frame $e_a{}^\mu$, normalized as $e_a{}^\mu e_b{}^\nu g_\mn=\eta_{ab}$, where $\eta_{ab}=\text{diag}(-1,1,1,1)$. Using $S^\mu=S^a e_a{}^\mu$ in Eq.~\eqref{eq:parallel transport}, one obtains
\begin{align}~\label{spin evolution local}
	v^\mu\pd_\mu S^{a}= -v^{\mu}\omega_{\mu}{}^a{}_b S^b\,,\qquad \omega_{\mu}{}^{a}{}_b =e^{a}{}_{\alpha} \nabla_{\mu} e_b{}^\alpha \,,
\end{align}
where $\omega_{\mu}{}^{a}{}_b $ is the spin connection associated to the tetrad. The  form~\eqref{spin evolution local} is not fully satisfactory for our purpose. On the one hand, neither side of~\eqref{spin evolution local} is covariant with respect to a general local Lorentz transformations $S^a\to \Lambda^a{}_b(x^\mu) S^b$. On the other hand, for a generic choice of the tetrad, the equation predicts a precession for the gyroscopes in the flat Minkowski background, since the spin connection is not necessarily vanishing in that case. It is yet possible to rewrite it in a way that resolves both issues and highlights its physical content. This may be achieved by adding $v^{\mu}\overline\omega_{\mu}{}^a{}_b S^b$ on both sides, where $\overline\omega_{\mu}{}^a{}_b$ is a background spin connection such that $\omega_{\mu}{}^a{}_b\to \overline\omega_{\mu}{}^a{}_b$ as the metric $g_\mn$ approaches a background metric $\eta_\mn$ (the flat metric in this setup). The resulting equation reads
\begin{align} \label{spin evolution local covariant}
	v^\mu\barD_\mu S^{a}= -v^{\mu}\left(\omega_{\mu}{}^a{}_b-\overline\omega_{\mu}{}^a{}_b\right) S^b\,,
\end{align}
where we have used that $\barD_\mu S^a \equiv \pd_\mu S^a+\overline{\omega}_{\mu}{}^a{}_{b} S^b$. Each side of~\eqref{spin evolution local covariant} is now covariant under arbitrary local Lorentz transformations, and the right-hand side vanishes over the background spacetime.  We can think of \eqref{spin evolution local covariant} as the time evolution of the spin with respect to a frame that is parallel transported with the background structure. 

To simplify further, we choose a comoving orthonormal frame $e_{a}{}^{\mu}=(v^\mu,e_{\hi}{}^{\mu})$, where the temporal basis vector coincides with the geodesic velocity $v^\mu$ of the gyroscope, while the spatial basis vectors, denoted by $e_{\hi}{}^{\mu}$ with $\hi\in\{1,2,3\}$, are not further specified. In this frame, the spin is purely spatial, $S^\mu=S^\hi e_\hi{}^\mu$, and~\eqref{spin evolution local covariant} reduces to
\begin{align} \label{eq:spin dynamics comoving}
	\dfrac{\dd S^\hi}{\dd\tau} =\Omega^\hi{}_\hj S^\hj\,,\qquad  \Omega^\hi{}_\hj =-v^{\mu}\big(\omega_{\mu}{}^\hi{}_{\hj}-\overline{\omega}_{\mu}{}^\hi{}_{\hj} \big)
\end{align}
where $\dd/\dd\tau \equiv v^\mu\barD_\mu$ and all frame indices are raised or lowered with the Kronecker delta $\de_{\hi\hj}$. Now, it turns out that the \textit{asymptotic flatness}, which we impose as a boundary condition, induces a natural background structure $\overline{\omega}_{\mu}{}^\hi{}_{\hj}$ uniquely fixed by the boundary metric on the celestial sphere~\cite{Seraj:2021rxd,Seraj:2022qyt}. In physical terms, the corresponding inertial frame is realized by ``distant stars'' on the celestial sphere. Therefore, \eqref{eq:spin dynamics comoving}~is interpreted as the time evolution of the spin with respect to a frame \textit{tied to distant stars}, and the antisymmetric tensor $\Omega^\hi{}_\hj$ is the precession rate in the $\hi\hj$ plane.

\subsection{Gyroscopes in asymptotically flat spacetimes} \label{sec: formal}

\subsubsection{Precession equations for asymptotic gyroscopes}

The metric far from a localized source of gravitational waves is suitably expressed in Bondi coordinates $(u,r,\th^\a)$ with $u$ the null retarded time, $r$ the areal distance from the source localized around $r=0$, and $\th^\a$ a coordinate system on the sphere. As an asymptotic expansion in $r$, it reads
\begin{align} \label{eq:metric Bondi}
	\dd s^2 =& -\left( 1 - \frac{2G\mu}{r c^2} + \cO\left( \frac{1}{r^2} \right) \right)\dd u^2 - 2 \left( 1 + \cO \left( \frac{1}{r} \right) \right) \dd u\,\dd r \nonumber \\ &  + r^2 \left(\gamma_{\a\b} + \frac{1}{r} \, C_{\a\b} + \cO \left( \frac{1}{r} \right)\right) \dd\th^\a \, \dd\th^\b + \left( \DD_\b C_\a{}^\b\dd u \, \dd\th^\a + \cO \left( \frac{1}{r} \right) \right) \,,
\end{align}
where $G$, $c$ are the Newton constant of gravitation and the speed of light, respectively. The metric $\gamma_{\a\b}$ on the unit celestial sphere (or its inverse $\gamma^{\a\b}$) allows raising (lowering) spherical indices. In other occurrences, it is used to define the covariant derivative $\DD_\a$ on the sphere and the Levi-Civita tensor $\veps_{\a\b}$. The Bondi shear $C_{\a\b}(u,\th^\a)$, which is symmetric and trace-free with respect to $\gamma_{\a\b}$, encodes the gravitational waves emitted by the source. The metric~\eqref{eq:metric Bondi} shows explicit leading-order deviations from the Minkowski metric, while neither the Bondi mass aspect $\mu(u,\th^\a)$ nor the ``subleading'' corrections~\cite{Godazgar:2018vmm, Grant:2021hga, Compere:2022zdz} are required for the purpose of this paper, since the leading gyroscopic effect is entirely determined in terms of the Bondi shear, as we will see below.

Let us now consider a gyroscope at large but finite distance $r$ from the source. We aim at computing its precession with respect to a frame pointing towards distant stars. The details of the construction of the tetrad is given in~\cite{Seraj:2021rxd,Seraj:2022qyt} (see also~\cite{Freidel:2021fxf,Godazgar:2022pbx,Seraj:2022qqj}). Decomposing spatial directions into radial and transverse $\hi=\hr,\ha$, it was found in~\cite{Seraj:2021rxd,Seraj:2022qyt} that the passage of gravitational waves by the gyroscope induces a dominant precession in the transverse plane given by 
\begin{align} \label{eq:precession rates}
	\Omega_{\ha\hb} =
	\frac{\veps_{\ha\hb}}{r^2}
	\hat{\Omega}+\cO \left( \frac{1}{r^3} \right) \,,\qquad \hat{\Omega}(u,\th^\a) = \underbrace{\frac{c}{4} \DD_\a \DD_\b \tlC^{\a\b}}_{\WS} \underbrace{-\frac{1}{8}\dot{C}_{\a\b}\tlC^{\a\b}}_{\WH} \,,
\end{align}
where $\tl{C}_{\a\b}\equiv \veps_{\c\a}C^\c{}_\b$ is the dual shear.  The other component $\Omega_{\hr\ha}$ is subleading at large distance, scaling as $\cO(r^{-3})$.

\subsubsection{Gyroscopic memory}

After the passage of the wave, GWs induce a net rotation in the orientation of the gyroscope in transverse plane, whose angle is simply obtained by the time integral of~\eqref{eq:precession rates},
\begin{align} \label{eq:gyroscopic memory}
	\Delta\Psi& = \frac{\Delta \hat{\Psi}}{r^2} + \cO\left( \frac{1}{r^3} \right) \,,\qquad \Delta \hat{\Psi} = \int_{u_0}^u \dd v \big(\WS(v) + \WH(v)\big) \,.
\end{align}
The precession rate~\eqref{eq:precession rates} consists of a linear term in the shear and a quadratic term. They are denoted by S and H labels respectively, which is justified in two ways. On the one hand, the time integral in~\eqref{eq:gyroscopic memory} kills all but the zero frequency mode in the Fourier expansion of the linear term, and is thus referred to as \textit{soft}, while the quadratic term includes gravitons of all frequencies, and is referred to as \textit{hard}. On the other hand, the first term coincides exactly with the \emph{spin memory} effect~\cite{Pasterski:2015tva,Nichols:2017rqr,Compere:2019gft,Mitman:2020pbt}, while the second term measures the total \emph{helicity}, the difference between the number of right-handed and left-handed gravitons at a given point on the celestial sphere~\cite{Seraj:2021rxd,Seraj:2022qyt,Maleknejad:2023nyh,Liu:2024rvz, Dong:2024ily,Kol:2022bsd,Hosseinzadeh:2018dkh}. 

Let us end this section by a remark on terminology. In the following, we will call $\WS$ the linear precession, as it is a linear function of the Bondi shear,  and $\WH$ the nonlinear precession, as it is quadratic in the shear. However, in the context of gravitational memory effects, there is a splitting of the memory into linear and nonlinear pieces~\cite{Christodoulou:1991cr,Blanchet:2013haa,Grant:2021hga,Nichols:2017rqr}, also called the ordinary and null memory effects \cite{Bieri:2013ada}. In the latter terminology, linearity (nonlinearity) instead refers to first (second) order effects in a perturbative solution of Einstein's equations. These two terminologies do not coincide in any way, since the Bondi shear itself contains perturbative effects of all orders. To avoid any confusion, we will never use the term linear or nonlinear memory.

\section{Multipole expansion of the precession rate} \label{sec: multipole expansion}

\subsection{Precession in the holomorphic basis}

Our main results, Eqs.~\eqref{eq:precession rates} and~\eqref{eq:gyroscopic memory}, involve tensorial expressions constructed out of the STF Bondi shear and its covariant derivatives with respect to the round metric $\gamma_{\a\b}$ on the sphere. Remarkably, STF tensors of any rank in two dimensions have only two 
independent degrees of freedom, combinable into a single complex scalar. This correspondence provides an elegant and practical formulation of the problem in terms of \textit{spin-weighted} functions, which we explain here and further in appendix~\ref{appendix:holomorphic basis}. This formalism was developed in the representation theory of the rotation group~\cite{gelfand1963representations}, and independently in the spin-coefficient formalism of Newman and Penrose~\cite{Newman:1966ub} and is widely used in the context of GW theory.

Given a real orthonormal basis $E_{\hat{1}}{}^\a,E_{\hat{2}}{}^\a$ on a two-dimensional Riemannian manifold, one can construct a pair of null vectors, consisting of  $m^\a =( E_{\hat{1}}{}^\a + \ii \,E_{\hat{2}}{}^\a)/\sqrt{2}$ and its complex conjugate $\barm^\a = (E_{\hat{1}}{}^\a - \ii \,E_{\hat{2}}{}^\a)/\sqrt{2}$.  By construction $m^\a m_\a = 0 =\barm^\a \barm_\a$ and $m^\a \barm_\a = 1$. In our setup, the relevant manifold is a unit round sphere, whose metric and volume form can be recovered from the real and imaginary part of the product $m_\a\barm_\b$
\begin{align} \label{eq:m barm}
  m_\a \barm_\b&=\frac{1}{2} (\gamma_{\a\b}-\ii\, \veps_{\a\b}) \,.
\end{align}
Moreover, the product of  $m_{\a_1 \cdots\, \a_s}\equiv m_{\a_1} \cdots\, m_{\a_s}$ and $\barm_{\a_1 \cdots\, \a_s}$ provide a complete basis for STF tensors of rank $s$. In particular, the Bondi shear takes the form
\begin{align} \label{eq:complex shear def}
  C_{\a\b}=m_{\a\b} \, C+\barm{}_{\a\b} \, \barC\,,\quad \text{where} \quad C = \barm{}^{\a\b} C_{\a\b}\,,
\end{align}
where $C, \barC$ have spin-weight $-2, +2$ respectively (see appendix~\ref{appendix:holomorphic basis} for the definition and further details).
Now, let us compute the precession rate~\eqref{eq:precession rates} in this basis. The linear part $\WS$ involves
\begin{align}
  \DD_\a \DD_\b \, \tlC^{\a\b} =& \,\veps^{\a\c} \, \gamma^{\b\d} \, \DD_\c \DD_\d \, C_{\a\b}  =  \ii \left(m^{\a\b} \barm^{\c\d}-\barm^{\a\b} m^{\c\d}\right)  \DD_\c \DD_\d \, C_{\a\b} \,. 
\end{align}
In deriving the last term, we have used \eqref{eq:m barm} to rewrite the metric and the Levi-Civita tensor in terms of the null dyad, and the STF property  $m^\c\barm^\d \, C_{\c\d}=0$ of the shear tensor to simplify the result. Using the eth derivative $\eth$, which acts on $C$ as
\begin{align}
	\eth C &= \barm^{\a} \barm^{\b} m^\c \DD_\c C_{\a \b} \,,
\end{align}
as well as its conjugate $\bareth$, defined similarly but with $m^\c$ replaced by its complex conjugate $\barm^\c$ (see Appendix~\ref{appendix:holomorphic basis}), we thus find  
\begin{align} \label{eq:OmegaS}
  \WS &= \frac{c}{4\ii}(\eth^2 C-\bareth^2 \barC)=\frac{c}{2}\, {\rm Im} (\eth^2 C)
\,. \end{align}
The nonlinear term $\WH$ in~\eqref{eq:precession rates} is also easily expressed as
\begin{align} \label{eq:precession rate-nonlinear}
  \WH &= -\frac{1}{8} \dot{C}_{\a\b} \tlC^{\a\b}  = \frac{\ii}{8} \left( m_{\a\b} \, \dot{C} + \barm_{\a\b} \, \dot{\barC} \right) \left( m^{\a\b} C - \barm^{\a\b} \barC \right)= \frac{1}{4} \operatorname{Im}( \dot{C}\barC )\,.
\end{align}
Therefore, the total precession rate is given by 
\begin{align} \label{eq:precession rate complex}
\hat{\Omega}= \frac{1}{4} \operatorname{Im}( 2c \,\eth^2 C + \dot{C} \barC ) \,.
\end{align}
Note in particular that the result is of spin-weight 0, implying that the precession rate is independent of the choice of frame, see Eq.~\eqref{eq:frame rotation} in appendix~\ref{appendix:holomorphic basis}.

\subsection{Multipole expansion}

By construction, the complex shear $C$ is of spin-weight $s=-2$ and thus can be multipole expanded in the basis of spin-weight $-2$ spherical harmonics as
\begin{align} \label{eq:multipole expansion shear}
  C(u,\th^\a) = \sum_{\ell\geqslant 2,\, m} C_{\ell m}(u) \, \Ylms{-2}{\ell m}(\th^\a) \,,
\end{align}
where we use the shorthand notation $\sum_{\ell\geqslant 2,\, m}=\sum_{\ell=2}^{+\infty} \sum_{m=-\ell}^\ell$.  The complex conjugate $\barC=\sum_{\ell\geqslant 2,\, m} \barC_{\ell m}\,\barYlms{-2}{\ell m} $ can be expanded, using the property $\barYlms{-s}{\ell m}=(-1)^{m+s}\Ylms{s}{\ell-m}$, as 
\begin{align}
  \barC &= \sum_{\ell \geqslant 2,\, m} C_{\ell m}^* \, \Ylms{2}{\ell m} \,,\qquad  C_{\ell m}^* = (-1)^m \, \barC_{\ell-m} \,.
\end{align}
The Bondi shear $C_{AB}$ can also be expanded in terms of parity-definite real scalars $U(u,\th^\a),V(u,\th^\a)$ with even and odd parity respectively as 
\begin{align} \label{eq:shear parity}
  C_{\a\b} &= \DD_{\langle \a} \DD_{\b\rangle } U + \veps^\c{}_{(\a} \DD_{\b)} \DD_{\c}\,V \,,
\end{align}
In this relation, angle brackets denote the symmetric trace-free part of the tensor under investigation. An equivalent expression for \eqref{eq:shear parity}, which is more democratic between even and odd-parity terms is
\begin{align} \label{eq:shear parity 2}
  C_{\a\b} &= \DD_{\langle\a} \DD_{\b\rangle} \, U + \veps^\c{}_{\a} \DD_{\langle \b} \DD_{\c \rangle} \,V \,.
\end{align}
Using~\eqref{eq:shear parity 2} in the second equation in~\eqref{eq:complex shear def}, we obtain
\begin{align} \label{eq:C vs cC}
  C=\bareth^2 \barZ\,,\qquad Z = U + \ii\, V \,.
\end{align}

Multipole expansion is an integral part of the post-Newtonian/multipolar post-Min\-kowskian formalism~\cite{Blanchet:2013haa}, which facilitates the derivation of the radiation field in terms of the source parameters. In our language, this corresponds to
\begin{align}\label{eq:cC multipole expansion}
  \begin{split}
	Z &= \sum_{\ell \geqslant 0}Z_{L} n^L\,,\qquad Z_L = \frac{4G}{c^{\ell+2} \ell!} \frac{1}{\ell(\ell-1)} \left( U_L- \frac{2\ell\, \ii}{c\, (\ell+1)} V_L \right)\,.
  \end{split}
\end{align}
where $U_{L},V_{L}$ are respectively the mass and current radiative multipoles in the basis of STF harmonics $n^L\equiv n^{\langle i_1}\cdots n^{i_\ell \rangle}$.
Alternatively, we can multipole expand in the basis of \textit{spherical} harmonics, by writing $Z = \sum_{\ell \geqslant 0}Z_{L} n^{L'}\de^{\langle L\rangle}_{\langle L'\rangle}$ and using the completeness relationship $\de^{\langle L\rangle}_{\langle L'\rangle}=4\pi \ell!/(2\ell+1)!!\,\sum_{m=-\ell}^{\ell}\cY^{\ell m}_L \bar{\cY}^{\ell m}_{L'}$.  The numerical coefficients $\cY^{\ell m}_L$  relate the two bases as $Y^{\ell m}(\th^\a)=\cY^{\ell m}_L n_L(\th^\a) $ \cite{Thorne:1980ru}. The result is
\begin{align} \label{eq:cC spherical}
  Z &= \sum_{l\geqslant 0\,, m} Z_{\ell m} Y^{\ell m}\,,\qquad Z_{\ell m} = \frac{4\pi \ell!}{(2\ell+1)!!} Z_{L}\,{\bar\cY}^{\ell m}_{L}\,.
\end{align}
Using~\eqref{eq:C vs cC} and~\eqref{eq:cC spherical}, the complex shear is expanded as 
\begin{subequations}\label{eq:complexified radiative multipoles}
\begin{align} 
  C & = \sum_{\ell\geqslant 2 \,, m} C_{\ell m}\, \Ylms{-2}{\ell m} & C_{\ell m} = \frac{G}{\sqrt{2}c^{\ell+2}} \left( {U}_{\ell m}-\ii \frac{V_{\ell m}}{c} \right) \,, \\
  \barC &= \sum_{\ell\geqslant \,2, m} C^*_{\ell m} \, \Ylms{2}{\ell m} & C^*_{\ell m} = \frac{G}{\sqrt{2}c^{\ell+2}} \left( {U}_{\ell m}+\ii \frac{V_{\ell m}}{c} \right) \,,
\end{align}
\end{subequations}
where the spherical multipoles $U_{\ell m},V_{\ell m}$ are related to STF multipoles $U_{L},V_{L}$ as~\cite{Thorne:1980ru}
\begin{subequations}
\begin{align}
  U_{\ell m}&=\frac{16\pi}{(2\ell+1)!!} \sqrt{\frac{(\ell+1)(\ell+2)}{2\ell(\ell-1)}} U_{L} \, \bar{\cY}_{L}^{\ell m} \,, \\
  V_{\ell m} &= \frac{-32\pi \ell}{(\ell+1)(2\ell+1)!!} \sqrt{\frac{(\ell+1)(\ell+2)}{2\ell(\ell-1)}} V_{L} \, \bar{\cY}_{L}^{\ell m}\,.
\end{align}
\end{subequations}

\subsection{Multipole expansion of the precession}

\subsubsection{Linear precession rate}

Expanding~\eqref{eq:OmegaS} in terms of radiative multipoles~\eqref{eq:complexified radiative multipoles}, and using the property~\eqref{eq:eth Y} of the $\eth$ operator, we find 
\begin{align} \label{eq:linear precession}
  \WS &= -\frac{G}{4c^2}\sum_{\ell \geqslant 2\,, m} \sqrt{\frac{(\ell+2)!}{2(\ell-2)!}} \, \frac{V_{\ell m}}{c^\ell}\,Y^{\ell m}\,, 
\end{align}
We observe that the linear precession is purely determined in terms of current multipoles.

\subsubsection{Nonlinear precession rate}

The nonlinear precession rate is proportional to the imaginary part of $\dot{C} \barC =$  \linebreak $ \sum_{\ell_1m_1} \sum_{\ell_2m_2} \dot{C}_{\ell_1 m_1}{C}^*_{\ell_2 m_2} \, \Ylms{-2}{\ell_1 m_1}\,\Ylms{2}{\ell_2 m_2}$. Being of spin-weight 0, the latter can be expanded in terms of ordinary spherical harmonics, using the orthogonality property of spin-weighted harmonics~\eqref{eq:orthogonality harmonics}. One finds
\begin{align}
  \dot{C} \barC &= \sum_{\ell_1\geqslant 2,\, m_1} \sum_{\ell_2 \geqslant 2,\, m_2} \sum_{\ell \geqslant 0 \,, m} \GG{\ell m}{\ell_1 m_1}{\ell_2 m_2} \,\dot{C}_{\ell_1 m_1} C^*_{\ell_2 m_2}Y^{\ell m} \,,
\end{align}
where
\begin{align}
  \GG{\ell m}{\ell_1m_1}{\ell_2m_2} &= (-1)^m \sqrt{\frac{(2\ell_1+1)(2\ell_2+1)(2\ell+1)}{4\pi}} \ThreeJ{\ell_1}{m_1}{\ell_2}{m_2}{\ell}{-m} \ThreeJ{\ell_1}{2}{\ell_2}{-2}{\ell}{0} \,.
\end{align}
Therefore, $\WH$ takes the form $ \WH = \sum_{\ell \geqslant 0\,, m} \hat{\Omega}^{\text{(H)}}_{\ell m} Y^{\ell m}$ with
\begin{align}
	\hat{\Omega}^{\text{(H)}}_{\ell m} &= \frac{1}{8\ii} \sum_{\ell_1\geqslant 2,\, m_1} \sum_{\ell_2\geqslant 2,\, m_2}\GG{\ell m}{\ell_1 m_1}{\ell_2 m_2} \, \Big(\dot{C}_{\ell_1m_1} C^*_{\ell_2 m_2} -C_{\ell_1 m_1} \dot{C}^*_{\ell_2 m_2} \Big)  \,.
\end{align}
Note that by the triangular property of the $3j$ symbols, the above is nonvanishing only for $|\ell_1-\ell_2|\leqslant \ell\leqslant \ell_1 + \ell_2$ and $|m|\leqslant \ell$. 
Now, using~\eqref{eq:complexified radiative multipoles} in the above result, we can express the hard precession rate in terms of radiative multipole moments as
\begin{align} \label{eq:nonlinear precession multipole expansion}
	\hat{\Omega}^{\text{(H)}}_{\ell m} (u)&= \frac{G^2}{16\ii\, c^4} \sum_{\ell_1,\, m_1}\sum_{\ell_2,\, m_2} c^{-(\ell_1 + \ell_2)}\, \GG{\ell m}{\ell_1 m_1}{\ell_2 m_2}  \nonumber\\
  &\qquad \times \Big[\;\big( \dot{U}_{\ell_1m_1} U_{\ell_2m_2} + \!\frac{1}{c^2} \dot{V}_{\ell_1m_1} V_{\ell_2m_2}\big) (1 - \!(-1)^{\ell_1+\ell_2+\ell}) \nonumber\\
    &\qquad\quad + \frac{\ii}{c} \big(\dot{U}_{\ell_1 m_1} V_{\ell_2 m_2} - \dot{V}_{\ell_1 m_1} {U}_{\ell_2 m_2}\big) (1 + \!(-1)^{\ell_1+\ell_2+\ell}) \Big] \,.
\end{align}
In this result, the time dependence is encoded in the radiative multipoles, and the post-Newtonian order is made explicit by the factors of $1/c$ in the result. The dominant effect is given by the $\dot{U}U$ term with $\ell_1=2=\ell_2$, since there exist odd multipolar orders $\ell=1,3$ for which the prefactor $1-(-1)^{\ell_1+\ell_2+\ell}$ is nonvanishing, and the triangular condition $0\leqslant \ell \leqslant 4$ is  satisfied.

An alternative expression for $\WH$, which makes the PN order more explicit, is obtained by permuting the sums in such a way that those over $\ell$ and $m$ become the outermost ones. The domain of variation of the dummy indices are to be modified accordingly. It is also convenient to denote $\ell'= \ell_1$ and eliminate $\ell_2$ in terms of the non-negative integer $p$ such $\ell_1 + \ell_2 + \ell = 2p$ if $\ell_1 + \ell_2 + \ell$ is even, and $\ell_1 + \ell_2 + \ell = 2p + 1$ if $\ell_1 + \ell_2 + \ell$ is odd. This yields
\begin{align}\label{eq: WH alternative}
 & \WH = \sum_{\ell \geqslant 0\,, m} \hat{\Omega}^{\text{(H)}}_{\ell m} Y^{\ell m}\,, ~ \text{with}\nonumber \\
  & \hat{\Omega}^{\text{(H)}}_{\ell m} = \frac{G^2}{8\ii \, c^7} \sum_{\ell' = 2}^{+\infty} \Big[ \sum_{p=0}^{\min(\ell,\ell')-1-\delta_{\ell\ell'}} \!\!   \frac{1}{c^{\ell+2(\ell'-2)-2p}}   \sum_{m'=-\ell'+\max(0,m-\ell+2p+1)}^{\ell+\min(0,m+\ell-2p-1)} \GG{\ell m}{\ell' m'}{\ell+\ell'-2p\, m-m'} \times \nonumber \\ & \qquad \qquad \qquad \qquad \qquad \qquad \qquad \qquad \times [\dot{U}_{\ell' m'} U_{\ell+\ell'-2p-1\, m-m'} + \frac{1}{c^2} \dot{V}_{\ell' m'} V_{\ell+\ell'-2p-1\, m-m'}] \nonumber \\ & \qquad \qquad \qquad \quad + \sum_{p=0}^{\min(\ell,\ell')-\delta_{|\ell-\ell'| \leqslant 1}} \!\!  \frac{\ii}{c^{\ell+2(\ell'-2)-2p+2}} \sum_{m'=-\ell'+\max(0,m-\ell+2p)}^{\ell+\min(0,m+\ell-2p)} \GG{\ell m}{\ell' m'}{\ell+\ell'-2p-1} \times  \\ & \qquad \qquad \qquad \qquad \qquad \qquad \qquad \qquad~ \times [\dot{U}_{\ell' m'} V_{\ell+\ell'-2p\, m-m'} - \dot{V}_{\ell' m'} U_{\ell+\ell'-2p\, m-m'}]  \,. \nonumber
\end{align}
From those two expressions of $\WH$, it is straightforward to figure out the leading post-Newtonian order of the various contributions to $\hat{\Omega}^{\text{(H)}}_{\ell m}$ for a given value of $\ell$ \ie, including terms proportional to the product of two mass multipole moments denoted as $UU$, terms made of the product of two current multipole moments, denoted as $VV$ and the mixed $UV$ contributions depending on both types of moments. The various leading orders are summarized in Table~\ref{tab:LO} and Table \ref{tab:multipolarity}.
\begin{table}[!h]
		\centering
		\begin{tabular}{|c|c|c|c|c|c|}
			\hline
			\diagbox{\makebox[10pt][l]{\hspace{0mm}{Type} }}{\makebox[15pt][l]{\hspace{2mm}\raisebox{-2.6mm}{$\ell$} }} & 0 & 1 & 2 & 3 & $\geqslant 4$ \\
			\hline
			$UU$ & --  & 8  & 9  & 8  & $\ell+5$\\ 
			\hline
			$VV$	& --   & 10 & 11 & 10 & $\ell+7$\\ 
			\hline
			 $UV$ & 9 & 10 & 9 & 10 & $\ell+5$\\
			\hline
		\end{tabular}
		\caption{This table shows the smallest power of $1/c$ that appears in each type of term contributing to $\hat{\Omega}^{\text{(H)}}_{\ell m}$ in Eqs.~\eqref{eq:nonlinear precession multipole expansion}, \eqref{eq: WH alternative}, for  given multipolarity $\ell$.}
		\label{tab:LO}
	\end{table}

It is also useful to know the set of multipolar orders that appear at a given post-Newtonian order. In Table~\ref{tab:multipolarity}, we display for each type of terms its leading post-Newtonian order, as well as the minimum and maximum multipolarities where it arises.
\begin{table}[!h]
  \begin{center}
    \begin{tabular}{|c||c|c|c|c|c|c|}
      \hline    & \multicolumn{3}{c|}{$n$ even, $\ell$ odd} & \multicolumn{3}{c|}{$n$ odd, $\ell$ even} \\
      \hline  \hline types of terms & $UU$ & $VV$ & $UV$ & $UU$ & $VV$ & $UV$\\
      \hline $n_\text{min}$ & 8 & 10 & 10 & 9 & 11 & 9 \\
      \hline $\ell_\text{min}$ & 1 & 1 & 1 & 2 & 2 & 0\\
      \hline $\ell_\text{max}$ & $n-5$ & $n-7$ & $n-5$ & $n-5$ & $n-7$ & $n-5$\\ \hline
    \end{tabular}
    \caption{In this table, we display the the smallest power of $1/c$ that appears in each type of term contributing to $\hat{\Omega}_{\text{(H)}}$ , as well as the minimum and maximum multipolarities  for each type of term.  see Eqs.~\eqref{eq:nonlinear precession multipole expansion}, \eqref{eq: WH alternative}.} \label{tab:multipolarity}
  \end{center}
\end{table}

\subsection{Total helicity flux}

We can compute the total helicity flux by integrating~\eqref{eq:nonlinear precession multipole expansion} over the sphere. We will see that this quantity vanishes for planar orbits. Therefore a net helicity flux is only possible for binary systems of spinning objects.

Integrating~\eqref{eq:nonlinear precession multipole expansion} over the sphere kills all spherical harmonics except $\ell=0$. The triangular property of the $3j$-symbols then implies that $\ell_1=\ell_2$. As a result, $\ell_1+\ell_2+\ell$ is even, which kills $UU, VV$ terms. Also, \scalebox{0.6}{${\displaystyle \ThreeJ{\ell}{m}{\ell}{-m}{0}{0}}$} $= (-1)^{\ell+m}/\sqrt{2\ell+1}$, which implies that $\GG{00}{2 m_1}{2 m_2} = (-1)^m/\sqrt{4\pi}$. Using these, we are left with
\begin{align}
  \int_{S^2} \dd^2\mathit{\Omega} \; \WH &= \frac{G^2}{8c^5}\sum_{l\geqslant 2\,,m} \frac{(-1)^m}{c^{2\ell}} \big( \dot{U}_{\ell m} V_{\ell\, -m} - \dot{V}_{\ell m} {U}_{\ell\, -m} \big) \,.
\end{align}
For planar orbits, it can be proven in general that~\cite{Faye:2012we}
\begin{align} \label{eq:planar radiative multipoles}
  \begin{split}
	U_{\ell m}&=0\,,\qquad \ell+m=\text{odd} \,, \\
	V_{\ell m}&=0\,,\qquad \ell+m=\text{even} \,.
  \end{split}
\end{align}
As a result, the total helicity flux vanishes for planar orbits, no matter the orbit is bound or unbound. We conclude that in order to have a total helicity flux from binary systems, we need to have spinning companions.

\section{Post-Newtonian sources} \label{sec: PN}

In this section, we will compute the gyroscopic memory sourced by simple binary systems. To this end, we will use available analytic expressions for $C_{\a\b}$~\cite{Blanchet2008}, which have been derived within the post-Newtonian/multipolar post-Minkowskian (PN/MPM) formalism~\cite{BlanchetLR_2024}. The following subsection provides some background material on the dynamics of quasi-circular binaries as well as the PN/MPM formalism and its application to those systems. For practical purposes, the reader can skip these discussions and jump to Eqs.~\eqref{eq:PN multipoles quasi circular}, displaying the leading radiative multipole moments of the source under consideration. This is essentially what we need for the rest of the paper.


\subsection{Quasi-circular binary system}

\subsubsection{Post-Newtonian dynamics of spinless compact binaries}

In Newtonian gravity, the dynamics of two pointlike masses $m_1,m_2$ is a solvable model, integrable in the sense of Liouville. In the center-of-mass frame, the problem reduces to the motion of a body of reduced mass $\mu=m_1m_2/M$ with $M=m_1+m_2$. Generic solutions are uniquely determined by three constants of motion: energy $E$, angular momentum $J_i$, and the direction of the Laplace-Runge-Lenz vector $A^i$ in the orbital plane.

At higher post-Newtonian orders, the Lagrangian for the \textit{conservative} dynamics of point masses is derived by integrating out the gravitational degrees of freedom. A clear distinction must be made between those referring to the near zone (surrounding the matter source but of size much smaller than the wavelength) and those referring to the exterior zone (excluding the matter source). The evaluation of the functional integral whose logarithm essentially gives the reduced action may be achieved either by means of the saddle point method, as in Fokker's approach, or, alternatively à la Feynman, by expanding the integral around a Gaussian kernel and applying Wick's theorem~\cite{Damour:2017ced} (see~\cite{BlanchetLR_2024, Porto:2016pyg, Schafer:2018jfw} for recent reviews of this type of computation). In the case of spinless objects, the integrability of the resulting Hamiltonian is preserved at higher post-Newtonian orders, in general relativity, still neglecting dissipative effects due to gravitational radiation (see \eg, \cite{Damour:2000kk} for a Hamiltonian analysis at the 3PN order). Bound spinless binaries, notably, follow planar quasi-elliptic trajectories subject to periastron advance. \emph{When taking radiation into account}, both the eccentricity and the separation monotonically decrease, so that, in the last stage of its history, the system becomes \emph{quasi-circular} and shrinks, up to the merger of the two companions.

In the constant plane of the motion, the binary orbits are then \emph{quasi-circular} and \emph{inspiraling}. They are most conveniently described in polar coordinates $\rho, \psi=\int \dd t\, \omega(t)$, which are not independent, but are related, instead, through a relativistic version of Kepler's law, $GM/\rho^3 = \omega^2 [1 + \cO(1/c^2)]$. The remainder $\cO(1/c^2)$ entering the right-hand side of this relation, in the post-Newtonian framework, takes the form of an asymptotic series in the so-called post-Newtonian parameter $x\equiv (GM\omega/c^3)^{2/3}$. This quantity is dimensionless, of order $1/c^2$, and is linked to the coordinate velocity $v^i$ through $x=v^2/c^2 + \cO(1/c^4)$. However, unlike $v^2$, the parameter $x$ is an adiabatic invariant~\cite{Arnold:1989who, Damour:2000kk}, defined independently of the particular post-Newtonian coordinates that are adopted to describe the dynamics.

Now, due to the shrinking of the orbits, the separation $\rho$ is not exactly constant, nor is the orbital frequency $\omega$, but they both depend adiabatically on time. Their dependence may be calculated from the energy flux-balance equation~\cite{LL, Thorne:1980ru}, implying that $\dot{\rho}=\cO(1/c^5)$, and, by virtue of post-Newtonian Kepler's law, $\dot{x}=\cO(1/c^5)$. More precisely, the evolution of $x=x(u)$ at leading order is given by~\cite{LL}
\begin{align}
\dot{x}&=\frac{64}{5}\frac{c^3}{GM^2}\mu x^5 \left[1+\cO(x)\right] \,.
\end{align}
The corrections $\cO(x)$ are currently known up to the 4PN order~\cite{blanchet:2023sbv}.
This approximation starts to break down past the innermost stable circular orbit (ISCO), after which, typically, numerical methods are required to describe the binary evolution.

\subsubsection{Multipolar post-Minkowskian formalism}

At high orders, the waveform and fluxes are computed using the MPM formalism in harmonic coordinates. It involves three main tasks~\cite{BlanchetLR_2024}: (i) Constructing  perturbatively the general solution to vacuum Einstein's equations outside the matter, using a post-Minkowskian expansion in formal powers of the gravitational constant $G$; the perturbation of each order is expressed as a multipole expansion, depending on six (yet) unspecified sets of multipole moments. (ii) Solving Einstein's equation in the near zone as a post-Newtonian expansion in formal powers of $1/c$ up to a homogeneous solution at each order. (iii) By means of asymptotic matching techniques, comparing the multipole expansion of the near-zone gravitational field with the near-zone expansion of the exterior field; this matching condition determines for once the full formal solution, including integral expressions for the multipole moments in terms of the near-zone stress-energy pseudo-tensor, and the homogeneous near-zone solutions in terms of functional integrals of the multipole moments.

To read off the waveform, one has to kill terms proportional to $\ln^q r/r$ (with $q$ being any positive integer) by moving to (asymptotically or exact) radiative coordinates~\cite{blanchet:1986dk,blanchet:2020ngx,blanchet:2023pce}. The radiative multipoles $U_L$ and $V_L$ are then obtained as functionals of the six MPM moments, whose explicit expressions are known in terms of the source, as explained earlier. For inspiraling binaries of spinless compact objects, they are given in terms of the PN parameter $x(u)$ and the phase $\phi(u)$ (see~\cite{BlanchetLR_2024} for details).

\subsubsection{Post-Newtonian radiative moments of compact binaries}

As this point, we consider a Cartesian-like grid $(t,x^i)$ in harmonic coordinates, with its origin at the center of mass of the binary and its $z$ axis aligned with the angular momentum. In the corresponding spherical coordinates, the observation point is given by $(\tilde{r},\th,\phi)$ while the reduced mass is specified  by $\big(\rho(t),\pi/2, \psi(t)\big)$. The shear in Bondi gauge, used in the previous sections, can be deduced  with the help of the method of~\cite{blanchet:2020ngx,blanchet:2023pce} from the result obtained by means of the aforementioned procedure from tasks (i) to (iii) . Up to $\cO(1/r)$ corrections, as well as a free BMS transformation, Bondi angular coordinates coincide with the harmonic spherical coordinates, while $u=t-\tilde{r} -2G M_\text{ADM}/c^2\log \tilde{r} +o(1)$ and $r= \tilde{r}+G M/c^2+o(1)$, with $M_\text{ADM}$ being the Arnowitt-Deser-Misner (ADM) mass of the system.

On the other hand, to fix the phase of the complex shear $C$ in Eq.~\eqref{eq:complex shear def}, we must choose a specific vector $m^\a$, or, equivalently, the polarization basis $E_{\hat{1}}{}^\a,E_{\hat{2}}{}^\a$ on the unit sphere. We will resort here to the usual coordinate basis of spherical coordinates $E_{\hat{1}}\equiv E_{\hat{1}}{}^\a \pd_\a= \partial_\th$ and  $E_{\hat{2}}{} =(\sin \th)^{-1}\partial_\phi$. With this convention,  $U_L$ and $V_L$ introduced in the multipole expansion~\eqref{eq:cC multipole expansion} are the same as in the post-Newtonian literature~\cite{Thorne:1980ru}. A similar statement holds for the spherical multipole moments $U_{\ell m}$ and $V_{\ell m}$.

The radiative multipoles required for the purpose of this paper then read
\begin{subequations} \label{eq:PN multipoles quasi circular}
\begin{align}
U_{22}&=-8 \sqrt{\frac{2 \pi}{5}} M  c^2 \, \nu \, x \ee^{-2 \ii\verythinspace \psi} \,,&	U_{20} &= \frac{4}{7} \sqrt{\frac{5 \pi}{3}} M  c^2 \,\nu\, x,   \\
V_{21} &= \frac{8}{3} \sqrt{\frac{2 \pi}{5}} \delta M  c^3 \,\nu \, x^{3 / 2} \ee^{-\ii\verythinspace \psi}, & V_{3,0} &= -\frac{32}{5}\sqrt{\frac{3\pi}{35}}Mc^4\nu^2 x^{7/2}\,.
\end{align}
\end{subequations}
where $\delta=(m_1-m_2)/M$, $\nu=\mu/M$. A complete list can be inferred from~\cite{Blanchet2008}\footnote{Note however that, due to a different choice of polarization vectors, the quantity $r(h_+ - \ii h_\times)$ in~\cite{Blanchet2008} differs from the complex shear $C$ as defined in the present paper by a minus sign (neglecting $\cO(1/r)$ corrections).}. We present the results in terms of the phase of the waveform defined as 
\begin{align}
\varphi(u) &= \psi(u)-\phi+\frac{\pi}{2}+\cO \left( \frac{1}{c^3} \right)\,.
\end{align}

\subsection{Post-Newtonian expansion of the precession rates}

\subsubsection{Leading effects}

The leading effect in~\eqref{eq:linear precession} is given by 
\begin{align} \label{eq:WS leading}
  \WS &= -\frac{G\sqrt{3}}{2c^{4}}\bigg[ \sum_{|m|\leqslant 2} V_{2 m}Y^{2m} + \sqrt{5} \sum_{|m|\leqslant 3} V_{3 m}Y^{3m} \bigg] + \cO\Big(\frac{1}{c^6}\Big) \,,
\end{align}
For the nonlinear precession, as clear from~\eqref{eq:nonlinear precession multipole expansion}, the leading effect originates from $\ell_1=\ell_2=2$. This entails a 4PN effect from $UU$ terms, a 4.5PN effect from $UV$ terms, and a 5PN effect from $VV$ terms. Focusing on the leading order contribution, we note that the factor $(1-(-1)^{\ell_1+\ell_2+\ell})$, implies that $\ell=1,3$, and thus
\begin{align} \label{eq:WH leading}
  \WH &= \frac{G^2}{16\ii\, c^8} \sum_{|m_1|\leqslant 2} \sum_{|m_2|\leqslant 2} \sum_{\ell=1,3} \GG{\ell m}{2 m_1}{2 m_2} Y^{\ell m} \dot{U}_{2 m_1} U_{2 m_2} + \cO\Big(\frac{1}{c^9}\Big) \,,
\end{align}
where $\GG{\ell m}{2 m_1}{2 m_2}$ is non zero for $m=m_1+m_2\leqslant \ell$ due to properties of $3j$-symbols.

\subsubsection{Axisymmetric mode}

In the previous results, the axisymmetric mode, \ie, $m=0$ is of special status. The reason is that once the result is matched to the source, it takes the form $A(t)\exp(\ii m\phi(t))$, in terms of an adiabatic amplitude $A(t)$ and a fast oscillating phase $m\phi(t)$. The axisymmetric case $m=0$ is special, since it a DC effect, which accumulates over time and builds the leading contribution in the gyroscopic memory~\eqref{eq:gyroscopic memory}. In fact, we will show that the time integral of an axisymmetric mode leads to a $c^5$ enhancement with respect to the integrand.
Restricting our attention to axisymmetric modes in \eqref{eq:WH leading} by setting $m=m_1+m_2=0$,  we find
\begin{align}
  \WH^\ax = \frac{5G^2}{16\ii \, c^8} \sum_{\ell=1,3} \sqrt{\frac{2\ell+1}{4\pi}} \ThreeJ{2}{2}{2}{-2}{\ell}{0} Y^{\ell 0} \sum_{|m_1| \leqslant 2} \! \ThreeJ{2}{m_1}{2}{-m_1}{\ell}{0} \dot{U}_{2 m_1} U_{2\, -m_1} + \cO\Big(\frac{1}{c^9}\Big) \,. \nonumber
\end{align}
In the sum over $m_1$, the $3j$ symbol involving $m_1=0$ vanishes since $\ell$ is odd. The positive and negative $m$ then combine into
\begin{align} \label{eq:precession ZM}
  \WH^\ax =& \frac{5G^2}{8c^8}\sum_{\ell=1,3} \frac{2\ell+1}{4\pi}
  \ThreeJ{2}{2}{2}{-2}{\ell}{0} \,P_{\ell}(\cos\th) \sum_{m_1=1,2} \! \ThreeJ{2}{m_1}{2}{-m_1}{\ell}{0}\; \mathrm{Im} \big( \dot{U}_{2 m_1} U_{2\, -m_1}\big) \nonumber \\
  & + \cO\Big(\frac{1}{c^9}\Big) \,,
\end{align}
where $P_{\ell}(x)$ is the associated Legendre polynomial.

\subsection{Precession and memory from quasi-circular binary systems}

For planar binary systems, the property~\eqref{eq:planar radiative multipoles} of radiative multipoles can be used to simplify the precession rates~\eqref{eq:WS leading} and~\eqref{eq:WH leading}. In the particular case of quasi-circular binary systems, the leading-order soft precession, as an expansion in powers of the PN parameter $x$ is is obtained by inserting \eqref{eq:PN multipoles quasi circular} in~\eqref{eq:WS leading}. The result is 
\begin{align} \label{eq:WS leading PN}
	\WS &= \frac{2GM\delta\nu}{c} x^{3/2} \sin\varphi \sin(2\th) + \alpha \sin 2\varphi \, x^2 +(\beta_1 \sin \varphi + \beta_3 \sin 3\varphi )x^{5/2} \nonumber \\ & + \frac{3}{5} \frac{GM\nu^2}{c}x^{7/2} \big(5\cos(3\th)+3\cos\th \big) + \cO(x^4)\,,
\end{align}
where $\alpha, \beta_1, \beta_3$ are known functions of $\theta$ and $\nu$, which are irrelevant for our leading order analysis, as will become clear below.
On the other hand, the leading term in the hard precession is 
\begin{align} \label{eq:WH leading PN}	
	\WH &= \frac{GM\nu^2}{c}x^{7/2}\bigg[\frac{5}{28}\cos(2\varphi) \big(\cos(3\th)-\cos\th\big) - \big(\cos(3\th)+7\cos\th\big)\bigg] + \cO(c^{-9})
\end{align}
We observe that the soft precession is dominant over the hard precession: the former is a 2PN effect($\sim c^{-4}$), while the latter is a 4PN effect. Moreover, we note that both effects contain fast modes $m\neq 0$, and adiabatic modes $m=0$. The former depends on time through both $x$ and $\varphi$, while the latter depends only on the slow variable $x$. Quite interestingly, the adiabatic modes in soft and hard precession rates are of the same order of magnitude, given in terms of $\Omega_0\equiv GM\nu^2 /c$, as
\begin{subequations}\label{eq:precession-axisymmetric}
\begin{align}
\WH^\ax &= -\Omega_0\, x^{7/2}\big(\cos(3\th)+7\cos\th\big)+\cO(c^{-9})\,,\\ 
\WS^\ax &= \frac{3}{5}\Omega_0\, x^{7/2}\big(5\cos(3\th)+3\cos\th\big)+\cO(c^{-9})\,,
\end{align}
\end{subequations}

\begin{figure}
	\centering
	\includegraphics[width=.8 \linewidth]{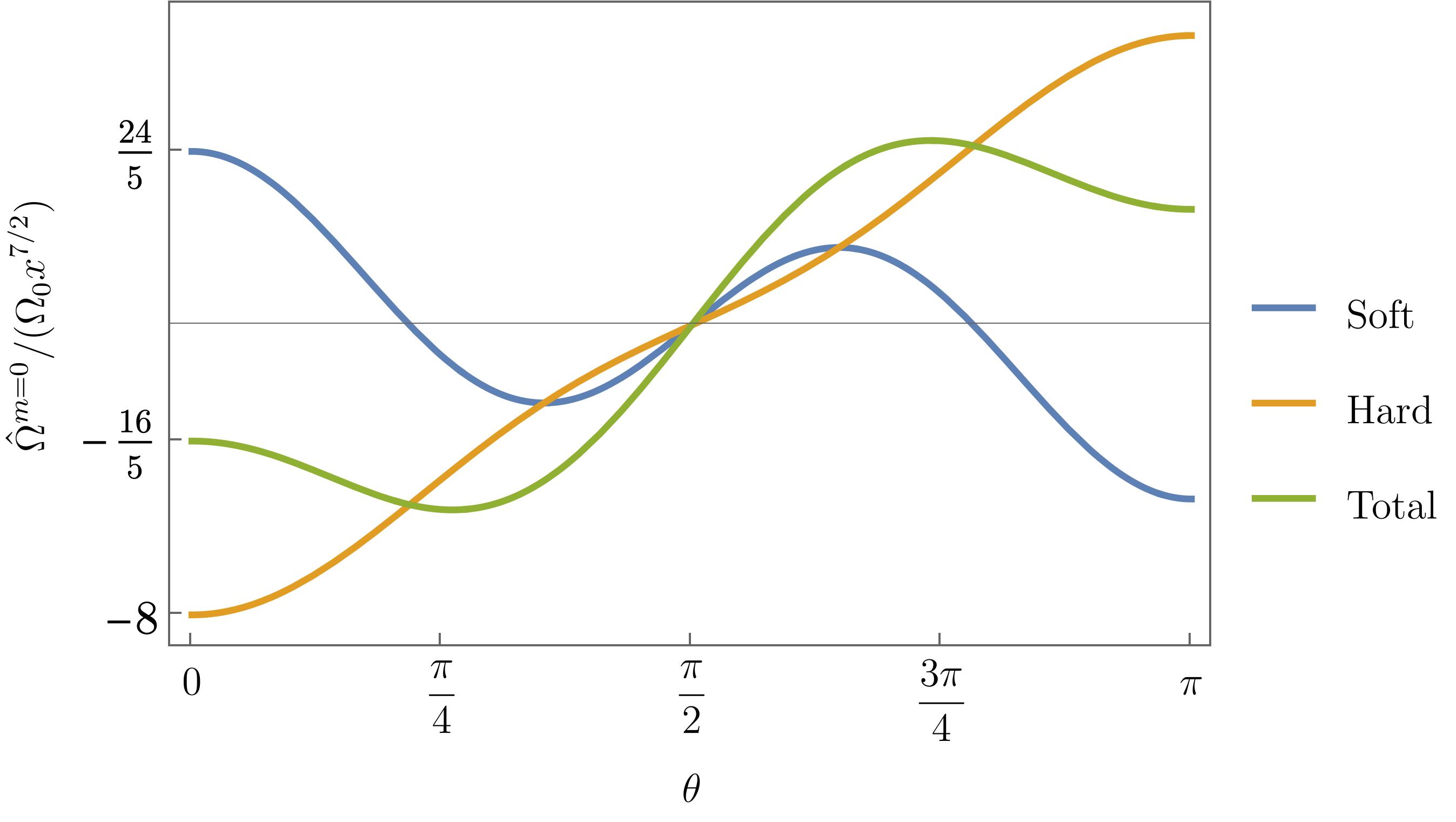}
	\caption{The leading axisymmetric contributions to the precession rate, given in \eqref{eq:precession-axisymmetric} as a function of the polar angle on the celestial sphere (with respect to the axis of rotation of the binary). The leading memory effects have similar angular dependence, see \eqref{eq:memory-axisymmetric}.} \label{fig:angular distribution}
\end{figure}

The leading memory effects are obtained by integrating precession rates over time. Such integrals are investigated in appendix~\ref{appendix:PN time integral}. The main results are Eqs.~\eqref{eq:time integral axisymmetric} and~\eqref{eq:time integral nonaxisymmetric}, indicating that there is a $c^5$ enhancement for axisymmetric (adiabatic) modes, while the time integration does not affect the PN order for non-axisymmetric modes. Therefore, while the leading term in the hard memory is the time integral of the leading term in \eqref{eq:WH leading PN}, the leading soft memory originates from a very subleading axisymmetric contribution, namely, the last term in \eqref{eq:WS leading PN}. The leading contribution to the memory effects \eqref{eq:gyroscopic memory} read 
\begin{subequations}\label{eq:memory-axisymmetric}
\begin{align} 
	\Delta\hat{\Psi}_{(\text{H})} &= \Psi_0\big(\cos(3\th)+7\cos\th\big) \left(\frac{1}{\sqrt{x}}-\frac{1}{\sqrt{x_0}} \right)+\cO(c^{-4})\,,\\
	\Delta\hat{\Psi}_{(\text{S})} &= -\frac{3}{5}\Psi_0\big(5\cos(3\th)+3\cos\th\big) \left(\frac{1}{\sqrt{x}}-\frac{1}{\sqrt{x_0}} \right)+\cO(c^{-4})\,. 
\end{align}	
\end{subequations}
The order of mangitude of the memory is given by $\Psi_0\equiv (5G^2M^2\nu)/(32c^4)$, which is 2PN. The leading gyroscopic memory is however enhanced to 1.5PN order due to the accumulation factor $(1/\sqrt{x}-1/\sqrt{x_0})$. The angular dependence of the celestial sphere is the same as~\eqref{eq:precession-axisymmetric}, depicted in Figure~\ref{fig:angular distribution}.

This equation shows the accumulated memory effect from an initial time $u_0$
at which $x(u_0)=x_0$, to the current time $u$ with $x= x(u)$. Given that the
quasi-circular model breaks down beyond the innermost stable circular orbit
(ISCO), one has to restrict to the regime where $x<x_\text{ISCO}=1/6$. On the
other hand, this model cannot be extended to the far past. Binaries are
formed over time, instead, through mechanisms such as dynamic capture, or the fragmentation of astrophysical clouds. Here, we must think of quasi-circular
evolutions as resulting from an initial value problem, where the binary is
circular at initial time $u_0$. Moreover, note that the Blanchet-Damour
formalism typically assumes a ``past-stationarity'' condition, \ie, that
gravitational fields are stationary before an initial time $u<u_0$. A more
careful analysis of IR divergences may be dealt with in a scattering setup,
see e.g.~\cite{Sahoo:2018lxl,Kehrberger:2023btg} for related works. In this
sense, $0<x_0\ll1$ can be thought of as an infrared cutoff.

\section{Concluding remarks} \label{sec: conclusion}

Let us briefly summarize the main results of this work and point out some future directions, which we plan to explore. It was showed in~\cite{Seraj:2022qyt} that gravitational waves from a binary system lead to a small change in the orientation of a distant gyroscope. There, the effect was estimated to be of order $G^2M^2/(r^2c^4)$, using dimensional analysis. We observe from Eq.~\eqref{eq:memory-axisymmetric} that dimensional analysis had correctly captured the leading result, except for the dimensionless accumulation factor $(1/\sqrt{x}-1/\sqrt{x_0})$ and the dimensionless symmetric mass ratio $\nu$. If the gyroscope is subject to long-lasting GW signal, the accumulation effect can lead to a large enhancement. At the same time, we note that the result is suppressed for extreme mass-ratio inspirals, for which $\nu\ll 1$. In any case, it is still very small overall, due to the $1/r^2$ factor, and thus unlikely to be observed in a realistic experiment, unless the gravitational radiation source turns out to be extremely close to the spinning body. 

On the other hand, in this paper, we have focused on the effect of GWs on a single gyroscope. Another scenario involves measuring the effect across a network of gyroscopes distributed on the celestial sphere. For example, we could think of pulsars surrounding a binary system at short distance, although this does not seem to be very realistic. In this case, we could study the correlation among the gyroscopes accordingly, and contrast it with the result~\eqref{eq:memory-axisymmetric}. 

Another interesting issue is to study more carefully the gyroscopic memory when the the test gyroscope is highly spinning. In this case, the parallel transport approximation fails, so that higher-order-in-spin effects must be taken into account, as outlined in Section~\ref{sec: Spin in GR}. These could be of relevance for the most rapidly spinning neutron stars, as probes of GWs.

\appendix

\section{Spin weighted functions on sphere} \label{appendix:holomorphic basis}

\subsection{Concept of spin weight, notations}

In this appendix, we recall how to define spin-weighted quantities associated to a null dyad on a two-dimensional sphere, while making explicit the notations and conventions used throughout the article. Our generic dyad will be denoted as $m^\a,\barm^\a$ and normalized in such as way that $ m^\a m_\a = 0 = \barm^\a \barm_\a$ and $ m^\a \barm_\a = 1$.  Inverting these relations allows finding the expression~\eqref{eq:m barm} of $m_\a \barm_\b$ in terms of the tensors $\gamma_{\a\b}$, $\veps_{\a\b}$. Under a $U(1)$ rotation in the tangent space of a given point on the sphere, the dyad transforms as
\begin{align} \label{eq:frame rotation}
	&  m^\a \to \ee^{\ii \verythinspace \th}m^\a\,,  \qquad \barm\to \ee^{-\ii \verythinspace \th}\,\barm \,.
\end{align}
For any tensor field $T$ of rank $p+q$, we may construct spin-weighted functions by contraction with basis vectors $m^\a$ and $\barm^\a$, \eg,
\begin{align} \label{eq:spin weighted function}
	T_s& \equiv T_{\a_1 \cdots \a_p \b_1 \cdots \b_q} m^{\a_1}\cdots m^{\a_p} \barm^{\b_1} \cdots \barm^{\b_q}
\end{align}
with $p$ factors of $m^\a$ and $q$ factors of $\barm^\a$. The complex function $T_s$ has spin-weight $s=p-q$, as it transforms under dyad rotations as $T_s\to \ee^{ \ii s\verythinspace \th} T_s$~\cite{Newman:1966ub}. We note that, using~\eqref{eq:m barm} in~\eqref{eq:spin weighted function}, mutual pairs of factors $m^\a \barm^\b$ may be traded for the metric and the epsilon tensor. Accordingly, the tensor $T$ in~\eqref{eq:spin weighted function} is reduced into its irreducible representations under the rotation group. In particular, a symmetric trace-free (STF) tensor $\hat{T}$ on the sphere has only two degrees of freedom, which are encoded in the single complex function $\mathcal{T}_s=\hat{T}_{\a_1\cdots \a_s}m^{\a_1}\cdots m^{\a_s}$. (The function $\mathcal{T}_{-s}$ of spin weight $-s$ obtained by contracting all indices with factors of $\barm^\b$ is not independent of $\mathcal{T}_s$, as they are complex conjugates.).

In this language, the covariant derivative acting on tensors on the sphere is replaced by the so-called eth $\eth$ derivative such that~\cite{Newman:1966ub}
\begin{align} \label{eq:eth def}
	\eth T_s &= m^{\a_1} \cdots m^{\a_p} \barm^{\b_1} \cdots \barm^{\b_q} m^\c \DD_\c T_{\a_1 \cdots \a_p \b_1 \cdots \b_q} \,,
\end{align}
while $\bareth T_s $ is defined by conjugating the vector that contracts with the derivative, \ie, $m^\c\to \barm^\c$ in the above equation. By construction, $\eth (\bareth)$ increases (decreases) the spin-weight by $+1$ $(-1)$. The normalization of $\eth$ in Eq.~\eqref{eq:eth def} differs from the one of Refs.~\cite{Newman:1966ub, Goldberg:1966uu}, for which there would be an extra factor $-\sqrt{2}$ in the right-hand side of~\eqref{eq:eth def}.

\subsection{Spin-weighted harmonics}

In this context, spin-weighted harmonics $\Ylms{s}{\ell m}$~\cite{Newman:1966ub, Goldberg:1966uu} provide a natural basis for spin-weighted functions on the sphere. They are defined in terms of standard harmonics $Y^{\ell m}\equiv \Ylms{0}{\ell m}$. For a positive integer $s$,
\begin{align} \label{eq: sYlm}
	\Ylms{s}{\ell m}&\equiv(-\sqrt{2})^s \sqrt{\frac{(\ell-s)!}{(\ell+s)!}} \;\eth^s Y^{\ell m} \,, \qquad	\Ylms{-s}{\ell m}\equiv (\sqrt{2})^s \sqrt{\frac{(\ell-s)!}{(\ell+s)!}} \;\bareth^s Y^{\ell m} \,,
\end{align}
and the spin-weighted harmonics vanish when $|s| > \ell$. The Condon-Shortley phase is chosen as in~\cite{Thorne:1980ru}, hence a factor $(-1)^m$ with respect to~\cite{Goldberg:1966uu}. The definition implies several important properties of spin-weighted harmonics. Under conjugation, its satisfies
\begin{align}
	\barYlms{s}{\ell m}=(-1)^{m+s}\Ylms{-s}{\ell-m} \,.
\end{align}
Moreover, for any non-negative $s\leqslant\ell$, the operators $\eth$ and $\bareth$ act as
\begin{align} \label{eq:eth Y}
		\eth\, \Ylms{s}{\ell m} &= -\sqrt{\frac{(\ell-s)(\ell+s+1)}{2}} \Ylms{s+1}{\ell m} \,, \quad	\bareth\, \Ylms{s}{\ell m} = \sqrt{\frac{(\ell+s)(\ell-s+1)}{2}} \Ylms{s-1}{\ell m} \,,
\end{align}
respectively. Finally, as a complete basis, the spin-weighted harmonics~\eqref{eq: sYlm} obey orthogonality relations given in terms of the $3j$ symbols:
\begin{subequations}
	\begin{align} \label{eq:orthogonality harmonics}
		&\int_{S^2} \dd^2\mathit{\Omega} \;\Ylms{s}{\ell_1 m_1}\,\barYlms{s}{\ell_2 m_2}=\de_{\ell_1\ell_2}\de_{m_1 m_2}\,, \\
		&\int_{S^2} \dd^2\mathit{\Omega} \;\Ylms{s_1}{\ell_1 m_1}\,\Ylms{s_2}{\ell_2 m_2} \, \Ylms{s}{\ell m} \nonumber \\ &
		\qquad \qquad = \sqrt{\frac{(2\ell_1+1)(2\ell_2+1)(2\ell+1)}{4\pi}} \ThreeJ{\ell_1}{m_1}{\ell_2}{m_2}{\ell}{m} \ThreeJ{\ell_1}{-s_1}{\ell_2}{-s_2}{\ell}{-s} \,,
	\end{align}
\end{subequations}
where $\dd^2\mathit{\Omega}$ denotes the element of two-dimensional solid angle.

\section{Time integration and memory effect} \label{appendix:PN time integral}

To compute the leading order contribution to the gyroscopic memory, we need to integrate the precession rates~\eqref{eq:WS leading PN},~\eqref{eq:WH leading PN} over time. These integrals take the form
\begin{align} \label{eq:time integral}
	I_{\alpha,m}(u) = \int_{-\infty}^{u} \dd t \,x(t)^{\alpha} \, \ee^{ \ii \verythinspace  m\verythinspace \varphi(t)} \,.
\end{align}
Note that $x(t)$ is an adiabatic variable, while $\phi(t)$ is a fast oscillatory variable. Therefore the case $m=0$ in~\eqref{eq:time integral} is a special case which reads
\begin{align} \label{eq:time integral axisymmetric intermediate}
	I_{\alpha,0}(u)= \int_{-\infty}^{u} \dd t \,x^{\alpha}=  \int_{-\infty}^{u} \dd x \,\frac{x^{\alpha}}{\dot{x}}
\end{align}
The time evolution of $x$ is obtained by the flux-balance equation for energy~\cite{LL}, implying that 
\begin{align} \label{eq:x evolution}
	\dot{x}&=\frac{64}{5}\frac{c^3\nu}{GM} \,x^5 \,\Big(1+\cO\Big(\frac{1}{c^2}\Big)\Big)
\end{align}
This equation enables the computation of axisymmetric mode integrals, when inserted into~\eqref{eq:time integral axisymmetric intermediate} for $m=0$, which yields
\begin{align} \label{eq:time integral axisymmetric}
	I_{\alpha,0}(u) &= \frac{5}{64} \frac{GM}{c^3\nu} \frac{1}{\alpha-4} x^{\alpha-4} \Big(1+\cO\Big(\frac{1}{c^2}\Big)\Big) \,,& \alpha &\neq 4 \,.
\end{align}
On the other hand, the case $m\neq 0$ can be processed by the following replacement in~\eqref{eq:time integral}
$$e^{\ii m \varphi}=\frac{1}{\ii m}\frac{\pd}{\pd \varphi}e^{\ii m \varphi}=\frac{1}{\ii m \omega}\frac{\pd}{\pd t}e^{\ii m \varphi}=\frac{GM}{c^3} \frac{x^{-3/2}}{\ii m }\frac{\pd}{\pd t}e^{\ii m \varphi}\,,$$
and performing integration by parts to move time derivatives from the fast variable to the slow variable. We thus find
\begin{align} \label{eq:Im intermediate}
	I_{\alpha,m}(u)&=\frac{GM}{c^3}\frac{1}{\ii m}\left[x^{\alpha-3/2}e^{\ii m \varphi}\Big\vert_{-\infty}^{u}-\int_{-\infty}^{u} \dd t  e^{\ii m \varphi} \frac{\dd }{\dd t}\big(x^{\alpha-3/2}\big)\right] \,.
\end{align}
Assuming that $\alpha >3/2$, the first term vanishes at the lower bound and the integral localizes to an instantaneous expression (this does not hold for $\alpha=3/2$ which we come back to later). Moreover, using~\eqref{eq:x evolution}, the integrand of the integral in~\eqref{eq:Im intermediate} is $\dd(x^{\alpha-3/2})/\dd t=(\alpha-3/2)(64/5)(c^3\nu)/(GM)\,x^{\alpha+5/2}$, and the integral takes the same form as~\eqref{eq:time integral}. We can therefore repeat the above algorithm, which implies
\begin{align} \label{eq:time integral nonaxisymmetric}
	I_{\alpha,m}(u)&=\frac{GM}{c^3} \frac{\ee^{ \ii m\verythinspace \varphi(u)}}{\ii m}x(u)^{\alpha-3/2} \left[1-(\alpha-3/2)\frac{64}{5}\frac{\nu}{\ii m} x^{5/2}+\cdots\right]\,, ~\alpha>\frac{3}{2},~m\neq 0\,.
\end{align}
We observe that the result is a perturbative PN expansion, where the first correction is 2.5PN subleading with respect to the leading term. Therefore, for the purpose of our work, only the leading term is sufficient. Subleading terms can be found by repeating the above procedure to reach the desired order.

For $\alpha\leqslant 4$ in~\eqref{eq:time integral axisymmetric} and $\alpha\leqslant 3/2$ in~\eqref{eq:time integral nonaxisymmetric}, the integrals are not convergent.  However, the quasi-circular orbit cannot be trusted in the far past, as we discussed below \eqref{eq:memory-axisymmetric}. Therefore, we regulate the integrals by modifying the lower bound from $-\infty$ to $u_0$, which can be a large but finite negative constant.

\section*{Acknowledgements}

AS thanks Adam Pound and Simone Speziale for useful discussions. GF thanks Geoffrey Compère and Université Libre de Bruxelles for supporting a visit related to this project. This research was supported in part by the Université Libre de Bruxelles and the Perimeter Institute for Theoretical Physics. Research at Perimeter Institute is supported by the Government of Canada through the Department of Innovation, Science and Economic Development and by the Province of Ontario through the Ministry of Colleges and Universities.  

\bibliography{refs.bib}

\end{document}